\documentclass[fleqn,10pt, twocolumn]{wlscirep}
\usepackage[utf8]{inputenc}
\usepackage[T1]{fontenc}
\usepackage{jabbrv}
\usepackage{graphicx}
\usepackage{epstopdf}
\usepackage{subfig}
\usepackage{float}
\usepackage{caption}
\usepackage{amsmath}
\usepackage[normalem]{ulem}

\newcommand{\Teff}{$T_{\rm eff}$}  
\newcommand{\kms}{km\,s$^{-1}$}
\def\ione{\,{\sc i}}
\def\ii{\,{\sc ii}}

\def\v{\,{\sc v}}

\title{Spectroscopic evidence for a large spot on the dimming Betelgeuse}

\author[1]{Sofya Alexeeva}
\author[1,2,*]{Gang Zhao}
\author[3,4]{Dong-Yang Gao}
\author[3]{Junju Du}
\author[5]{Aigen Li}
\author[3]{Kai Li}
\author[3]{Shaoming Hu}
 
\affil[1]{CAS Key Laboratory of Optical Astronomy, National Astronomical Observatories, Chinese Academy of Sciences, Beijing, 100101, China}
\affil[2]{School of Astronomy and Space Science, University of Chinese Academy of Sciences, Beijing, 100049, China}
\affil[3]{Shandong Key Laboratory of Optical Astronomy and Solar-Terrestrial Environment, School of Space Science and Physics, Institute of Space Sciences, Shandong University, Weihai, Shandong, 264209, China}
\affil[4]{School of Astronomy and Space Science, Key Laboratory of Ministry of Education, Nanjing University, Nanjing, 210046, China}
\affil[5]{Department of Physics and Astronomy, University of Missouri, Columbia, MO 65211, USA}
\affil[*]{Corresponding author. E-mail: gzhao@nao.cas.cn}

\begin{abstract}
{\bf During October 2019 and March 2020, the luminous red supergiant Betelgeuse  demonstrated an unusually deep minimum of its brightness. It became fainter by more than one magnitude and this is the most significant dimming observed in the recent decades. While the reason for the dimming is debated, pre-phase of supernova explosion, obscuring dust, or changes in the photosphere of the star were suggested scenarios. Here, we present spectroscopic studies of Betelgeuse using high-resolution and high signal-to-noise ratio near-infrared spectra obtained at Weihai Observatory on four epochs in 2020 covering the phases of during and after dimming. We show that the dimming episode is caused by the dropping of its effective temperature by at least 170 K on 2020 January 31, that can be attributed to the emergence of a large dark spot on the surface of the star. }
\end{abstract}
\begin{document}

\flushbottom
\maketitle
\thispagestyle{empty}

\noindent 

\section*{Introduction}

\label{sec:intro}
Betelgeuse ($\alpha$ Orionis) is one of the closest luminous red supergiants (RSGs) to the Earth and due to its brightness it can be seen by naked-eye in the night sky. 
Betelgeuse appears to be on the late evolutionary stage of massive stars
and sooner or later it will explode as a Type II-P Supernova and turn into a relativistic compact object. It is promising to offer us an amazing celestial show. Such events occur rarely in human era therefore at present Betelgeuse attracts not only professional astronomers, but also has been widely covered in mass media. For astronomers, it is an important astronomical laboratory for investigation of stellar evolution, stellar atmosphere, pre-supernova explosion, and dust condensation.

Betelgeuse is classified as a semi-regular variable star with spectral type M2 Iab. 
During the period from October 2019 to mid March 2020, Betelgeuse demonstrated a deep minimum of its brightness. 
On 2019 October 25, its visual $V$ magnitude was +0.679 mag (Universal Time (UT) 11.06). After that, it was losing its luminosity with an average rate of 0.01 magnitudes per day, reaching the minimum on 2020 February 9 ($V$ = +1.638, UT 23.33) and it is the deepest one within the past 8 years.  
However, already in March 2020, Betelgeuse was rapidly recovering luminosity at an average rate of 0.02 magnitudes per day and on 2020 April 19, it became even brighter than the day before dimming started ($V$ = +0.487, UT 02.57). One of the interesting features of this great dimming is that the rate of brightening of Betelgeuse ($\sim$0.02 mag/day) is two times higher than the rate of its dimming ($\sim$0.01 mag/day).
This re-brightening event has also led to intense discussions in science community. 
To explain this minimum of its brightness, several explanations have been proposed: pre-phase of supernova explosion, obscuring dust, or changes in the photosphere of the star.

The European Southern Observatory's (ESO) Very Large Telescope (VLT) reported about the unusual dimming of Betelgeuse. The images of the star's surface not only show it losing luminosity but also its shape seemingly changing (ref. \cite{ESO_press_releas}).
They presented the comparison image of Betelgeuse before and after its dimming. The observations, taken with the Spectro-Polarimetric High-contrast Exoplanet REsearch (SPHERE) instrument on ESO VLT in January and December 2019, show how much the star has faded and how its apparent shape has changed. 

The spectrophotometric study of Betelgeuse reported in ref. \cite{2020ApJ...891L..37L} suggested that the effective temperature (\Teff) on 2020 February 15 was \Teff\ = 3600$\pm$25~K and the derived small drop $\Delta$\Teff\ = 50 K was insufficient to explain the observed dimming. Since RSGs like Betelgeuse are surrounded by a gaseous and dusty circumstellar envelope created by their mass loss, which spreads heavy elements into the interstellar medium, it was proposed that the recent dimming was caused by screening from circumstellar dust.  The possible chemical composition of the dust and
the normalized masses of the dust grains that could condense around the cooling envelope of Betelgeuse and screen the light was predicted in ref. \cite{2020MNRAS.496L.122G}.

During September, October, and November 2019, three months before the dimming event, the ultraviolet spectroscopic observations from the Hubble Space Telescope (HST) / Space Telescope Imaging Spectrograph (STIS) revealed an increase in the ultraviolet spectrum and Mg\ii\ line emission from the chromosphere of the Betelgeuse \cite{2020ApJ...899...68D}. 
The detected variations in the Mg\ii\ k-line profiles gave evidence of the movement of material outwards. This movement, likely due to convective photospheric elements, was further aided with outward motions during the $\sim$400 day pulsation cycle. A relationship between the large convective cells and material movement was inferred via ultraviolet observations and suggested that the ejected material cooled as a dust cloud and likely caused the exceptional optical dimming of Betelgeuse in February 2020 \cite{2020ApJ...899...68D}. 

 However, only small changes in the gas velocities were detected from the high spectral-resolution observations of [Fe\ii\,] 25.99 $\mu$m and [S\ione\,] 25.25 $\mu$m emission lines from Betelgeuse obtained with the Stratospheric Observatory for Infrared Astronomy (SOFIA) Echelon Cross Echelle Spectrograph (EXES) during the dimming event on February 2020. It was argued that the dimming could not be due to dust given the lack of variation in the outflow velocities and line ratios \cite{2020ApJ...893L..23H}.

The TiO (titanium oxide) and near-infrared (near-IR) photometric data obtained at the Wasatonic Observatory \cite{2020ApJ...905...34H} found a significantly cooler mean effective temperature of Betelgeuse during the dimming event than that inferred from the spectrophotometric observations reported in ref. \cite{2020ApJ...891L..37L}. 

 Submillimetre (sub-mm) observations from the James Clerk Maxwell Telescope (JCMT) and Atacama Pathfinder Experiment (APE) over a period of 13 years which includes the dimming period, indicates a decline in sub-mm flux by 20$\%$ compared to values before the dimming event \cite{2020ApJ...897L...9D}. The modelling presented in ref. \cite{2020ApJ...897L...9D} indicates that the condensed dust cannot be the reason of this sub-mm flux drop, and only changes in the photosphere of a star (change in radius or temperature) might lead to the sub-mm flux reduction. 

 Alternative scenario was presented by George et. al \cite{2020A&A...640L..21G}, who suggested that the dimming of Betelgeuse might be caused due to critical transition in the pulsation dynamics.

Here, we present the spectroscopic study of Betelgeuse with the high-resolution and high signal-to-noise ratio (S/N) spectra obtained at Weihai Observatory of Shandong University on 2020 January 31, March 19, April 4, and April 6. We determine the effective temperature (\Teff), surface gravity (log~$g$), and microturbulence ($V_{mic}$) of Betelgeuse by employing a Markov Chain Monte Carlo (MCMC) forward-modeling method. We rule out pulsations as a possible mechanism of temperature decreasing and confirm that this dimming episode is caused by the dropping of its effective temperature by at least 170 K on 2020 January 31, that can be attributed to an emergence of a large dark spot on the surface of the star. We deduce the chemical compositions of C, N, O, Na, Mg, Ca, Ti, Cr, Fe, and Sr at the pre-dimming, dimming, and post-dimming episodes and show that the chemical abundances are stable in time if variable temperature is considered. Our results are consistent with ESO-VLT \cite{ESO_press_releas}, ultraviolet HST-STIS \cite{2020ApJ...899...68D}, infrared SOFIA-EXES \cite{2020ApJ...893L..23H}, and JCMT submillimeter \cite{2020ApJ...897L...9D} observations that the dimming of Betelgeuse is unlikely caused by dust obscuration.

\section*{Results} \label{sec:Result}

\subsection*{Effective temperature.} The physical parameters of Betelgeuse in the pre-dimming, dimming and post-dimming epochs are presented in {\bf Table 1} and {\bf Figure 1, a -- d}. We have found that on 14-02-2012, when the visual magnitude $V$ = 0.426, the effective temperature of Betelgeuse was \Teff\ = 3632$\pm$7~K. During the dimming on 31-01-2020, when $V$ = 1.609, the effective temperature was found to be lower by 156~K compared to pre-dimming episode. After 31-01-2020, the \Teff\ was increasing and became \Teff\ = 3534$\pm$3~K on 19-03-2020, then \Teff\ = 3611$\pm$7~K on 04-04-2020, and finally reached \Teff\ = 3646$\pm$9~K on 06-04-2020, which is close to its pre-dimming temperature. The effective temperature was found to be well correlated with the visual magnitude at all epochs of observations ({\bf Fig.~1a} and {\bf Fig.~1c}). 
 
 The changes in \Teff\ are sufficient to explain the recent variations in $V$ magnitude. To demonstrate this, we adopted the formula for the $V$ band bolometric correction, which was computed by employing MARCS stellar atmosphere models \cite{2006ApJ...645.1102L, 1975A&A....42..407G, 2003ASPC..288..331G, 1992A&AS...94..527P, 2003ASPC..298..189P}:

\begin{eqnarray*}
BC_V = -298.954+217.532\frac{T_{eff}}{1000 (K)} \\ 
  - 53.14\left(\frac{T_{eff}}{1000 (K)}\right)^2+4.34602\left(\frac{T_{eff}}{1000 (K)}\right)^3. && (1)
\end{eqnarray*}  
 
According to (1), a rise of 170 K in \Teff\  from 3476~K of 31-01-2020 to 3646~K of 06-04-2020 would correspond to a decrease in the $V$ magnitude by $\Delta V$ $\approx$ 0.753 mag. This is fairly sufficient to account for the observed brightening of $\Delta V$ = 0.918$\pm$0.171 mag, within the uncertainty. Similarly, a rise in \Teff\ from 3476 K of 31-01-2020 to 3534 K of 03-19-2020 would lead to a decrease of $\Delta V$ $\approx$ 0.301 mag, which, within the uncertainty, is also fairly enough to explain the observed brightening of $\Delta V$ = 0.456$\pm$0.186 mag, and a rise in \Teff\ from 3534 K of 19-03-2020 to 3646 K of 06-04-2020 would cause a decrease of $\Delta V$ $\approx$ 0.452 mag. This closely agrees with the observed brightening of $\Delta V$ = 0.453$\pm$0.162 mag.

\subsection*{ Evidence for a large cool spot.}
What could lead to a temperature drop by 170~K?
It should be noted that the lowest temperature obtained on 31-01-2020, \Teff\ = 3476~K, 
would be one of the coolest effective temperatures measured for a Galactic RSG. Although unusual, an effective temperature as low as 3365$\pm$134~K was found for the red supergiant UY Sct \cite{2013A&A...554A..76A}. Among 74 Galactic RSGs, only one star cooler than 3500~K was found \cite{2005ApJ...628..973L}. 
According to the stellar evolution theory, the Hayashi limit (the lower limit of \Teff\ for stars in hydrostatic equilibrium) for a star like Betelgeuse is $\sim$3500~K
(see, e.g., Figure 6.5 in \cite{2017ars..book.....L}). 
Nevertheless, the significance of the Hayashi limit for RSGs is still open to debate. It has been suggested that the stars moved to the forbidden region of the 
 Hertzsprung-Russell (H-R) diagram are not able to maintain their hydrostatic equilibrium and this may actually drive episodes of enhanced mass loss. 
However, there are also some individual stars, which still have a \Teff\ that places them to the right of the Hayashi limit, see e.g. \cite{2017ars..book.....L}.
It is still unclear whether these stars are in fact in hydrostatic equilibrium (and their observational effective temperatures were misinterpreted), or truly challenge evolutionary theory with unexpected cold temperatures, or these stars have cool spots on the surface that decrease their average temperature. 
 
We believe that it is unlikely that the entire Betelgeuse became cooler by 170~K. 
It could be caused by a dark star-spot on the surface of Betelgeuse. The presence of spots on the surface of Betelgeuse and other RSGs is a well known phenomenon, see e.g. \cite{1992MNRAS.257..369W, 1997MNRAS.291..819W, 2000MNRAS.315..635Y, 2009A&ARv..17..251S, 2018A&A...614A..12M}. These spots are likely a consequence of convective flows or cool convective cells, which are widely believed to be present in such stars \cite{1975ApJ...195..137S, 1986ApJ...307..261D, 1990MNRAS.245P...7B, 1997MNRAS.285..529T}. The size of spots can probably be large. For example, $H$-band interferometric observations of Betelgeuse, showed two spots at the surface of the star and one of the spots is as large as one half of the stellar radius \cite{2009A&A...508..923H}. 

We present an interpretation of temperature deviation as a changing of surface area of dark spots. 
The ESO-VLT observations \cite{ESO_press_releas} presented the images of the Betegeuse's surface before and after its dimming in January and December 2019 [\url{https://www.eso.org/public/images/eso2003c/}]. It is clearly seen, that about a half of the star has faded significantly. 
The dark region seen in the optical VLT observations could be explained by a large spot, whose temperature is lower than the typical temperature of Betelgeuse before the dimming. If we assume that in the minimum of brightness, on 31-01-2020, 50$\%$ of stellar surface was covered with dark spots, the temperature of spots would be 3306~K. The dark-spot covered surface would change from 50$\%$ on 31-01-2020 to 10$\%$ on 04-04-2020. The scheme of stellar surface and the best fits of narrow spectral region 7705 -- 7811~\AA\, for four epochs are presented in {\bf Figure~2, a -- d}. 

Taking into account its slow rotation velocity of $V$~sin$i$ = 5.47~\kms, the inclination of the polar axis on the line of sight $i$ = 60$^{\circ}$ and a rotation period, $P$/sin$i$, of 36 years \cite{2018A&A...609A..67K}, the star would rotate only by $\sim$~2$^{\circ}$ around its axis from 31-01-2020 to 06-04-2020. Such a small angle would not lead to significant change of surface area due to rotation of star. 
We did not find any significant changes in surface gravity at all epochs of observations from 31-01-2020 to 06-04-2020, log~$g$ remains constant within the error bars with a mean value log~$g$ = 0.22$\pm$0.04 ({\bf Fig.~1b}). Although on 14-02-2012, the log~$g$ is slightly lower (by 0.11~dex), but still consistent within 3$\sigma$.
If we assume a change in acceleration due to gravity as a result of a change in radius, $R$, (according to $g$ = GM/$R^2$), and take a 3$\sigma$ limit on the change of log~$g$ = 3*0.04 = 0.12, we can conclude that the $R$ could change at most 12$\%$ during the period of brightness recovering from 31-01-2020 to 06-04-2020. 

Even a small change in radius ($\sim$10$\%$) at constant \Teff\ would lead to a change in luminosity, $L$. If we know definitely that \Teff\ decreased by $\sim$170~K and this decreasing is enough to explain changing in luminosity, it means that the radius of star was definitely not changing much, because $L\sim$ $T_{\rm eff}^4R^2$. Otherwise, if both \Teff\ and $R$ are decreasing simultaneously, it would lead to a deeper dimming. Another case, if \Teff\ is falling down and $R$ is increasing simultaneously, then $L$ would not change at all and we would not observe any dimming (or we would observe brightening instead of dimming). Following these arguments we can rule out pulsations as a possible reason of the dimming of Betelgeuse.

The microturbulent velocity was found constant during all the epochs and was estimated 3~\kms. We did not find any significant differences in the broadening of the lines, that can indicate a constant the rotational velocity ($V$~sin$i$), which was adopted to be 5~\kms, according to \cite{1998AJ....116.2501U}. We have found overabundance in Fe relative to solar value at all epochs with average [Fe/H] = 0.09$\pm$0.03~dex 
({\bf Fig.~1d)}. 

\subsection*{ Chemical composition.}
We now elaborate the chemical composition of Betelgeuse with three different scenarios. 
The first scenario (\Teff\ = var, {\bf Figures~3, 4}) is that the effective temperature was changing according to our results presented in {\bf Table 1}. According to this scenario, the chemical abundances of all considered elements, C, N, O, Na, Mg, Ca, Ti, Cr, Fe, Sr remain constant within the error bars. This is reasonable since we do not expect the abundance of these elements 
change within such a short time span of $\sim$ 2 months. 
In {\bf Figure~3a}, carbon abundances were obtained under condition that N and O keep their constant values, A(N) = 8.6 and A(O) = 8.8. 
In {\bf Figure~3b}, nitrogen abundances were obtained at the condition that C and O keep their constant values, A(C) = 8.43 and A(O) = 8.8. 
In {\bf Figure~3c}, oxygen abundances were obtained at the condition that C and N remain constant values, A(C) = 8.43 and A(N) = 8.6. 
In {\bf Figure~4}, we present the chemical abundances of the remaining elements. 
Everywhere, the solar abundances are shown as a horizontal dashed line. 

The second scenario ({\bf Figures~3, 4}, open squares) is the effective temperature was constant (\Teff\ = 3646~K) and log~$g$ = 0.2 was adopted. At the constant value of \Teff\ at all five epochs, we are not able to reach the constant abundances of C, N, O, Mg at these epochs.
For example, {\bf in Figure~3a}, on 31-01-2020 (Julian date, JD - 2450000 = 8879.6 d), we would have A(C) = 8.13 (under condition \Teff\ = 3646~K, A(N) = 8.6, and A(O) = 8.8), that is by 0.28~dex lower than at the epoch of 06-04-2020 (JD - 2450000 = 8945.6 d). 
Another example, {\bf in Figure~3b}, on 31-01-2020 we would have A(N) = 7.99 (under condition \Teff\ = 3646~K, A(C) = 8.43, and A(O) = 8.8), that is by 0.61~dex lower than that at the epoch of 06-04-2020. 
In {\bf Figure~3c}, we would have A(O) = 9.12 (under condition \Teff\ = 3646~K, A(C) = 8.43, and A(N) = 8.6), that is by 0.32~dex higher than at the epoch of 06-04-2020. 
 Apparently, this is unlikely since it is difficult to imagine how the abundances of C, N, and O could considerably vary within at the period of only $\sim$ 2 months.

At the third scenario, we adopted the lowest temperature as constant value during all epochs (\Teff\ = 3476~K, {\bf Figures~3, 4}, open diamonds) and log~$g$ = 0.2. 
Analogically to the second scenario, we are not able to reach the constant abundances of C, N, O, Mg at these epochs at the constant temperature.
Here, we did not intend to analyze the absolute values of chemical abundances. We demonstrate that only if we assume variable \Teff\,, we can reach consistency in chemical composition of C, N, O, and Mg during all epochs of observations. As for other chemical elements, Na, Ca, Ti, Cr, Fe, Sr, it is impossible to conclude which scenario is more favorable, since the error bars are high.
 
{\bf Figure~5, a -- d} presents the molecular CN (cyanide) line at 10179~\AA, which was used along with others CN lines ({\bf Supplementary Table 1}) for the estimation of C, N, O conditional abundances. 
The observed molecular CN line at 10179~\AA\ is stronger on 14-02-2012 compared to 31-01-2020 {\bf (Figure~5a)}.
{\bf In Figure~5b}, the comparison of CN lines observed on 14-02-2012 and 19-03-2020, and the difference between two epochs is clearly seen. 
This CN molecular line is also stronger on 19-03-2020 compared to 04-04-2020 {\bf (Figure~5c)}. 
{\bf Figure~5d} presents a comparison between the CN from 2012 and the most recent spectrum on 06-04-2020. 
Although the spectral resolution of 14-02-2012 is slightly higher ($\it R$ = 65\,000) than that of other epochs ($\it R$ = 40\,000), 
however it still allows to see the significant difference in line strength ({\bf Figure~5, a -- d}). Together with observations as shown by dots, we present the best theoretical fits calculated with parameters specified on each panel according to the obtained values in {\bf Table 1}. If we assume constant chemical composition of C, N, O during all epochs, the difference in the strength of line can be interpreted by different temperature. 
Everywhere, carbon, nitrogen, and oxygen abundances are assumed to have constant values 8.43, 8.6, and 8.8 dex, respectively.

\subsection*{ Comparison with other studies.} 
Levesque $\&$ Massey \cite{2020ApJ...891L..37L} claimed that the obtained small drop, $\Delta$\Teff\ = 50 K, is insufficient to explain Betelgeuse recent optical dimming and they proposed that episodic mass loss and an increase in the amount of circumstellar dust along the sightline to Betelgeuse is the most likely explanation for its recent photometric evolution.
We have obtained different results. For example, at the dimming episode on 31-01-2020, our \Teff\ = 3476$\pm$4~K that is by 124~K lower compared to \Teff\ = 3600$\pm$25~K, obtained by ref. \cite{2020ApJ...891L..37L} on 2020 February 15. 
Although both studies are based on temperature sensitive TiO molecular lines, there are some differences in the methods.
Levesque $\&$ Massey \cite{2020ApJ...891L..37L} used the conventional, well-tested methodology, which was presented earlier in \cite{2005ApJ...628..973L, 2006ApJ...645.1102L} and tested on 74 Galactic RSGs. 
We have found that our technique provides a good agreement with \cite{2005ApJ...628..973L} for 7 common RSGs with a range of effective temperatures from 3440~K ($\alpha$Her) to 3815~K (HD~216946). We conclude that the differences in the methods could not explain the different results for Betelgeuse in the dimming episode. 

The ESO-VLT observations \cite{ESO_press_releas} provided the images of the Betegeuse's surface before and after its dimming in January and December 2019. 
It is clearly seen, that the star has faded and also its shape was seemingly changing indicating that something happened in the photosphere of the star or in its surroundings. The dark region seen in the optical VLT observations could be explained by a large spot, whose temperature is lower than the typical temperature of Betelgeuse before the dimming. Our findings are in line with these ESO-VLT observations, since the derived temperature drop could be caused by this large spot. We conclude that the radius of the star, R, would not change more than 12$\%$ during the period of brightness recovering by assuming a 3$\sigma$ limit on the change of log~$g$ = 0.12. This is in line with the ESO-VLT observation \cite{ESO_press_releas}, who demonstrated little or no radial change in the star. 

The sub-mm observations from the JCMT and APE showed that sub-mm flux was declined by 20$\%$ compared to values before the dimming event \cite{2020ApJ...897L...9D}.
The authors concluded that the dust envelope could not lead to the sub-mm flux reduction and only changes in the photosphere of a star (change in radius or temperature) might be responsible for this reduction. If the radius is supposed to be constant, the temperature drop is 200 K, which could be due to appearance of the star-spots with \Teff\ = 3250 K cover 50$\%$ of the visible surface or spots with \Teff\ = 3350 K cover 70$\%$ of the visible surface. These star-spots scenarios would also be accompanied by a $\sim$0.9 mag dimming in the $V$-band, and are qualitatively similar to the spectrum of Betelgeuse in ref. \cite{2020ApJ...891L..37L}. 
We found a similar temperature drop (170~K) in our study based on high resolution near-IR observations of Betelgeuse. 

The potential changes in the circumstellar flow during the dimming in 2020 February was investigated \cite{2020ApJ...893L..23H} by employing SOFIA-EXES high spectral-resolution observations of [Fe\ii]~25.99~$\mu$m and [S\ione]~25.25~$\mu$m emission lines from Betelgeuse. They detected very small changes in the gas velocities in the circumstellar flow,  that may indicate the dimming could not be due to dust. 

Recently, \cite{2020ApJ...905...34H} addressed the TiO variation based on photometry and did detailed temperature change calculations by comparing to synthetic photometry from the MARCS model photospheres and spectra. In both studies, ref. \cite{2020ApJ...905...34H} and this work, the TiO molecular absorption lines, which are strongly sensitive to temperature were employed, red/near-IR spectral ranges were investigated, and MARCS model atmospheres were adopted for comparison with theory. They employed photometric observations with three band wing red/near-IR filters with central wavelengths at 7190, 7540, 10240~\AA\,, while we used the high-resolution spectral analysis at 7000--11\,000~\AA\ wavelength range. According to ref. \cite{2020ApJ...905...34H}, \Teff\ declined from 3645$\pm$15 K (2019 September 21, 7 days bin) to 3520$\pm$25 K (2020 February 15, 22, and 29, 7 days bins), i.e., $\Delta$\Teff\ $\simeq$ $-$125 K, that is in line with our result. In {\bf Supplementary Figure 1}, we illustrate \Teff\ data derived from the TiO index in ref. \cite{2020ApJ...905...34H} and our work. 
Both studies agree with each other well and provide a full picture of the temperature dropping during the period of the dimming. 
Since ref. \cite{2020ApJ...905...34H} adopted an oxygen abundance by 0.28 dex lower compared to our study, it might lead to a systematic shift in temperature by $\sim$+20$-$30~K compared to our \Teff\ (the higher oxygen abundance in the atmosphere, the stronger molecular TiO lines employed as temperature indicators).

 The detected variations in the Mg\ii\ k-line profiles from the ultraviolet HST/STIS observations suggested movement of material outwards \cite{2020ApJ...899...68D}.
It was assumed that this outflow of material from the star was due to convective photospheric elements. 
These convective elements could caused the observed temperature drop, which was presented in ref. \cite{2020ApJ...897L...9D} and our study.

\section*{Discussion}\label{sec:Discussion}

During the period from October 2019 to mid March 2020, Betelgeuse  became fainter by more than one magnitude and it is the most significant dimming observed in the recent decades. 
We were motivated to search for possible explanations of this fading and carried out spectroscopic studies of Betelgeuse at the near-IR spectral region with the high-resolution and high-S/N spectra of Weihai Observatory of Shandong University obtained on 2020 January 31, March 19, April 4 and April 6 and the Observatory of the Canada France Hawaii Telescope (CFHT) obtained on 2012 February 14. 

We present a method to deduce \Teff\ and log~$g$, see Methods and the Effective temperature and Surface gravity subsections. The method is based on two (almost independent) indicators: the forest of 2365 TiO lines in the near-IR range at 7700 -- 7900~\AA\,, which is highly sensitive to \Teff\, and almost not sensitive to the gravity,
and the wings of Ca~\ii\ lines at 8542~\AA\ and 
8662~\AA\,, which are good indicator for gravity and slightly sensitive to temperature. 
Usually, the spectrophotometry and low resolution spectra ($\it R$ $\sim$ 1000) of broad TiO bands are preferred, see e.g. \cite{2017ars..book.....L}. Both methods are based on the TiO lines, which are highly sensitive to temperature, however instead of low resolution, we analyzed high resolution ($\it R$ $\ge$ 40\,000) spectra in the near-IR region.
We believe that the advantage of our method is the employment of a MCMC forward-modeling for fitting the data, where many different ways of spectral normalization are considered. It avoids the normalization by eye and helps to find the best normalization. 

We show that the effective temperature of Betelgeuse was dropping at least by 170~K during the period of its great dimming. 
 
From January to April 2020, the effective temperature was recovering to its original value.
For example, at post-dimming epoch on 06-04-2020 \Teff\ = 3646$\pm$9~K, while \Teff\ was 3632$\pm$7~K at pre-dimming epoch on 14-02-2012.
The rate of recovering correlated with the $V$ magnitude well.

We believe it is unlikely that the whole star became cooler by 170~K and attribute it to the emergence of a dark large spot on the surface on Betelgeuse. If we assume that the dark spot covers 50$\%$ of the visible surface of the star on 31-01-2020, the temperature of dark spot could be by 
340~K lower than the usual temperature (\Teff\ = 3646~K). 
The similar suggestion was also presented by 
\cite{2020ApJ...897L...9D}, who found a similar temperature drop 200~K in the minimum of brightness and reconciled it by star spots.

We did not find any significant changes in surface gravity at all epochs of observations from 31-01-2020 to 06-04-2020 and log~$g$ keeps their constant values within the error bars with a mean value log~$g$ = 0.22$\pm$0.04.  According to $g$ = GM/$R^2$, and taking a 3$\sigma$ limit on the change of log~$g$ = 3*0.04 = 0.12, we conclude that the radius of star was constant within 12$\%$ during the period of brightness recovering from 31-01-2020 to 06-04-2020.

Using two independent indicators to estimate \Teff\ and log~$g$, we rule out pulsation as a possible mechanism of temperature decreasing.
Since we know that \Teff\ was changing by 170~K (independent of log~$g$) and it is enough to explain change in $V$ $\sim$ 0.918, it is unlikely that the radius of star was changing, according to $L\sim$ $T_{\rm eff}^4R^2$. Otherwise, even a small change in radius would lead to change in luminosity and we would have much deeper dimming (if  \Teff\ decreases by 170~K and $R$ decreases simultaneously), no change in brightness (if \Teff\ decreases by 170~K and $R$ increases by 10$\%$), or even brightening of star (if \Teff\ decreases by 170~K and $R$ increases by more than 10$\%$ simultaneously).

We obtained the chemical compositions of Betelgeuse at the pre-dimming, dimming, and post-dimming episodes based on near-IR spectral region, where normalization can be done reliably 
({\bf Supplementary Table 2}). The developed line list was tested with solar spectrum.
The abundances for C, N, O, Na, Mg, Ca, Ti, Cr, Fe, and Sr would remain constant during the dimming episode
if variable temperature is considered. 
 In contrast, if we assume a constant temperature (the highest \Teff\ = 3646~K or the lowest one 3476~K) at the period of changing magnitude, we would face difficulty in explaining the changing of these chemical abundances in a short period of time.

\section*{Methods}

\subsection*{ Observations and data reduction} The observation list of Betelgeuse is presented in {\bf Supplementary Table 3}. 
High resolution ($\it R$ $\sim$ 40\,000) echelle spectra of Betelgeuse with wavelength range of 3710 -- 11000~\AA\ were obtained
on 2020 January 31, March 19, April 4, and April 6 with the bench-mounted, stabilized, fiber-fed Weihai Echelle Spectrograph (WES) \cite{2016PASP..128l5002G} attached to the 1 m telescope at the Weihai Observatory of Shandong University \cite{2014RAA....14..719H}, located at 122$^\circ$02$^\prime$58.6E and 37$^\circ$32$^\prime$09.3N. The telescope has an f/8 classic Cassegerain design and equatorial fork mount mechanical structure. The spectrograph covers 3710 to 11000~\AA\ in 107 spectral orders over the rectangular CCD.
The exposure time ($t_{exp}$) was 1200~s on January 31 and March 19
and 480~s on April 4 and 6 due to the increasing of brightness.
The S/N ratio measured at the continuum near 9000 \AA\ ranges from 80 in April to 500 in March 19.
The reduction of the CCD images was carried out in the context of the Image Reduction and Analysis Facility IRAF (IRAF is distributed by the National Optical Astronomy Observatories, which are operated by the Association of Universities for Research in Astronomy under cooperative agreement with the National Science Foundation), including bias subtraction, flat fielding,
background subtraction, wavelength calibration, continuum normalization, and flux calibration. The observations cover the epoch of the dimming and the subsequent recovery of brightness. On 2020 January 31, the $V$ magnitude was 1.609 mag, then Betelgeuse became slightly brighter with $V$ = 1.144 mag on March 19, and further recovery with $V$ = 0.754 mag and $V$ = 0.691 mag on April 4, and 6, respectively.  Each $V$ magnitude was calculated by averaging over all available magnitudes in the American Association of Variable Star Observers (AAVSO) database in a particular date of observations. The error for each $V$ magnitude is a standard deviation.

We used the high resolution ($\it R$ = 65\,000) spectra of Betelgeuse from the visible to the near-IR ranges obtained on 2012 February 14 with the Echelle Spectro Polarimetric Device for the Observation of Stars (ESPaDOnS) \cite{2006ASPC..358..362D} attached to the 3.6 m telescope of the CFHT Observatory located on the summit of Mauna Kea, Hawaii [\url{http://www.cfht.hawaii.edu/Instruments/Spectroscopy/Espadons/}]. Observations with this spectrograph cover the wavelength region from 3690~\AA\ to 10480 \AA\ and exposure time 0.8~s. Calibrated intensity spectral data were extracted from the ESPaDOnS archive through Canadian Astronomical Data Centre (CADC). The S/N measured at the continuum near 6000~\AA\ was estimated as 800 for Betelgeuse.

\subsection*{Model atmospheres and spectrum calculations.} Spherical local thermodynamic equilibrium  (LTE) model atmospheres MARCS \cite{2008A&A...486..951G} were used in our calculations. 
Extensive line lists for M-type stars were extracted via the facilities of the VALD3 [\url{http://vald.astro.uu.se/~vald3/php/vald.php}] database \cite{2015PhyS...90e4005R}. 
The synthetic line profiles are calculated by the code \textsc{synthV\_NLTE} \cite{2016MNRAS.456.1221R}.
The departure coefficients, $b_{\rm{i}}$ = $n_{\rm{NLTE}}$ / $n_{\rm{LTE}}$,
were computed with the code \textsc{detail} \cite{detail} based on the method of accelerated $\Lambda$ iteration \cite{rh91}. Here, $n_{\rm{NLTE}}$ and $n_{\rm{LTE}}$ are the statistical equilibrium and thermal (Saha Boltzmann) number densities, respectively. 
The \textsc{binmag} v.6 code [\url{http://www.astro.uu.se/~oleg/binmag.html}] \cite{binmag6} was used for automatic spectral line fitting and comparison with the observed spectrum. Throughout this study, the element abundances are determined from line profile fitting.

The Sun was adopted as benchmark star for testing all lines in near-IR region. We used the
MARCS model atmosphere 5777/4.44/0 and a depth-independent microturbulence of 0.9~\kms. The solar flux observations were taken from the Kitt Peak Solar Atlas
\cite{1984sfat.book.....K}. Our synthetic flux profiles were convolved with a profile that combines a rotational broadening of 1.8~\kms and broadening by macroturbulence with a
radial-tangential profile. For different lines, the most probable macroturbulence velocity, $V_{mac}$, was varied between 2 and 4~\kms. Hereafter, the element abundances are given in the scale, where for hydrogen A(H) = 12.

\subsection*{ Effective temperature.} Typically, the low resolution ($\it R$ $\sim$ 1000) spectrophotometry is used to determine the effective temperatures of RSGs. 
The absolute line strengths of TiO features at $\lambda\lambda$ 4761, 4954, 5167, 5448, 5847, 6158, 6658, and 7054~\AA\, are measured and
they allow us to avoid the spectral normalization procedure. 
The same TiO features are also employed to define the spectral types of M stars, e.g. see \cite{2017ars..book.....L}.

Here, for temperature determination, we have developed an alternative method based on a forest of TiO lines in near-IR range at 7700 -- 7900~\AA\,. Totally, 2365 TiO lines can be found in this range. 
The forest of molecular TiO lines in 7700 -- 7900~\AA\ is highly sensitive to temperature and only slightly to gravity.
We set up the log~$g$ to be constant at all epochs of observations and equal to 0.0. 
The theoretical spectra in the region of 7700 -- 7900~\AA\ were calculated with $V_{mic}$ = 3 \kms, $V$~sin$i$ = 5 \kms, $V_{mac}$ = 16 \kms, and the carbon, nitrogen, oxygen, calcium and titanium abundances are A(C) = 8.43, A(N) = 8.6, A(O) = 8.8 \cite{1984ApJ...284..223L}, 
A(Ca) = 6.34, and A(Ti) = 4.95. 
The dissociation energy of $D_0$ (TiO) = 6.87 eV \cite{1997CPL...266..335N}. The convolution with an instrumental profile was done according to the spectral resolution in {\bf Supplementary Table 3}. The model spectra were calculated with a step of 50 K for temperature, within this range, the linear interpolation was applied.

Since the lines are blended with each other, it complicates the finding of the exact level of continuum. The problem of spectral normalization procedure was solved by including the coefficient of re-normalization (Norm) as variable parameter in Monte Carlo simulations. It allows to simulate many different ways of normalization and its impact on the obtained temperature. Although, Norm does not play any physical role and was only considered for the accuracy of \Teff\ and log~$g$ determination.

We fit the forest of TiO lines in 7700 -- 7900~\AA\ to infer the effective temperature using an MCMC forward-modeling method built on the \textsc{emcee} package \cite{emcee}. The uncertainty is taken as the difference between the 84th and 50th percentile as the upper limit, and the difference between the 50th and 16th percentile as the lower limit for all model parameters. 
If the posterior distributions follow normal (Gaussian) distributions
then this equates to the 1$\sigma$ uncertainty for each parameter. The corner plots (\Teff\ vs. Norm) at four epochs of observations on 31-01-2020, 19-03-2020, 04-04-2020, and 06-04-2020 are presented in 
{\bf Figure~6} and in {\bf Supplementary Figure 2a} at 14-02-2012. The best fits of TiO bands for all investigated epochs of observations are presented in {\bf Figure 7}, where the 2365 TiO lines are marked by vertical lines on the panel (a).

\subsection*{Surface gravity.} To estimate the surface gravity, we used the same method as for the effective temperature. However, instead of the forest of TiO lines, we adopted the wings of Ca~\ii\ lines at 8542~\AA\ and 8662~\AA\,, which are good indicator for gravity and almost not sensitive to temperature. The model spectra were calculated with the determined \Teff\ for each epoch of observations. We take a step of log~$g$ of 0.2 dex, and within this range, the linear interpolation was applied. 
The Ca~\ii\ lines at 8542~\AA\ and 8662~\AA\ are not a subject of the non-local thermodynamic equilibrium (NLTE) effects (Sitnova, priv. comm.) and can be used in LTE assumption. 
The corner plots (log~$g$ vs. Norm) at four epochs of observations on 31-01-2020, 19-03-2020, 04-04-2020, and 06-04-2020 are presented in {\bf Supplementary Figure 3}, and at 14-02-2012, in {\bf Supplementary Figure 2b}.

Since the Ca~\ii\ lines at 8542~\AA\ and 8662~\AA\ are blended with plenty of other lines, we created a mask, which consists of some parts of the spectrum. 
These parts are presented in {\bf Supplementary Figure~4a, b} for two epochs of observations: 
{\bf (a)} panel presents Ca~\ii\ lines at the epoch of 14-02-2012 and {\bf (b)} panel presents Ca~\ii\ lines at 8542~\AA\ and 8662~\AA\ at the epoch of 31-01-2020. 
Only these spectral parts were taken for fitting, the result of which is presented by corner plots {\bf Supplementary Figures~3 and 2b}. 
In {\bf Supplementary Figure~4c}, we demonstrate the fitting of the wings of Ca~\ii\ lines at 8542~\AA\ and 8662~\AA\ in solar spectrum with the model ones. It allows to check atomic data such as van der Waals broadening constants, oscillator strengths to be sure that we are able to fit the wings of Ca~\ii\ lines at 8542~\AA\ and 8662~\AA\ properly.

\subsection*{Chemical composition.} The advantage of this research is that we used spectral observations at the near-IR region (8000 -- 11000~\AA) to derive the chemical composition. The spectral continuum can be determined reliably at this wavelength range. 
Nevertheless, for example, in blue region (4000 -- 6000~\AA), it is hard to find the location of continuum level, since there are plenty of atomic lines and molecular bands blended each other. The chemical composition was derived in LTE assumption except for Na and Mg,
for which the deviations from LTE were considered. 
The line list was developed according to the near-IR spectral region and in the first place it was tested on solar spectrum before applying to the Betelgeuse. 
Using this line list, we obtained chemical abundances from the Sun and then compared it with Betelgeuse.

The microturbulent velocity, $V_{mic}$ = 3 \kms, was obtained from Fe\ione\ lines listed in {\bf Supplementary Table 4}. The synthetic profiles
were convolved with a profile that combines a rotational broadening
of 5~\kms and broadening by macroturbulence with a radial-tangential profile. Macroturbulence is introduced to reproduce the line profile shape, and it does not affect the derived element abundance, if the line profile is fitted well.
For different lines of Fe~\ione, the macroturbulence velocity, $V_{mac}$, was varied between 10 and 15~\kms.
 {\bf Supplementary Figure~4d} demonstrates the iron abundances vs. the excitation energy ($E_{\rm exc}$) of the lower level for the investigated Fe~\ione\ lines for two epochs of observations: 14-02-2012 and 31-01-2020. In both epochs, there is no trend for A(Fe) vs. $E_{\rm exc}$ and the mean values of iron abundances are A(Fe) = 7.51$\pm$0.11 and A(Fe) = 7.53$\pm$0.15. Iron abundances from individual lines at all epochs of observations are presented in {\bf Supplementary Table 5}. Using the same line list, we obtained A(Fe) = 7.41$\pm$0.09 for the Sun.

Below, we briefly describe the method of chemical abundance determination for particular element. 

To determine C, N, O abundances, we employed CN molecular lines presented in {\bf Supplementary Table 1}. 
The best fits of CN molecular line 10179~\AA\, are presented in {\bf Figure~5}. 
The dissociation energy is $D_0$(CN) = 7.65 eV  \cite{1988ApJ...332..531B}. 
The CN lines depend not only on C and N abundances, but also on O abundance in the atmosphere \cite{1984ApJ...284..223L}. 
As a result, all presented C, N, O abundances are conditional ({\bf Figure~3}). 
For example, when we measure C, we keep A(N) = 8.6 and A(O) = 8.8. When we measure N, we keep A(C) = 8.43 and A(O) = 8.8.
Since, we were not able to get independent values of C, N and O abundances,
we adopted the initial values of them from \cite{1984ApJ...284..223L}.
We have found a set of values of carbon, nitrogen, and oxygen abundances A(C) = 8.43, A(N) = 8.6, A(O) = 8.8, at which 
we can get a good agreement between observational CN molecular lines and theoretical profiles.

We employed only one line of Na~\ione\ at 9961~\AA\,. The NLTE calculations with the model atom adopted from \cite{2014AstL...40..406A} showed that LTE abundances can be overestimated by $\sim$0.20~dex in the atmosphere with \Teff\ from 3476 to 3646~K and log~$g$ = 0.2. 
However, in solar atmosphere, the NLTE correction is only 0.02~dex for Na~\ione\ line at 9961~\AA\,. 
Since we employed only one line, the error bars were estimated from uncertainties in continuum level by 1$\%$.
We obtained that Na abundance in the atmosphere of Betelgeuse is much higher by about 0.60~dex relative to solar value, where the solar value was obtained from the same line in solar spectrum.  

We used the wings of the strong Mg\ione\ line at 8806.75~\AA\ and the line at 10312.53~\AA\,.
Although we provide NLTE values for Mg, our NLTE calculations with the model atom adopted from \cite{2018ApJ...866..153A} showed very small NLTE effects in Mg\ione\ line at 10312.53~\AA\, and the wings of Mg\ione\ line at 8806.75~\AA\, are not suffering from NLTE at all. We have found slightly overabundance ([Mg/H]~$\sim$~0.11) of Mg relative to solar value at all epochs {\bf (Figure~4b)}.  

By employing two Ca\ione\ lines at 6471.66~\AA\, and 10273.68~\AA\, we have found that Ca abundances are close to solar values within the error bars at all epochs of observations. 

We measured five lines of Ti\ione\ at $\lambda\lambda$ 10034.49, 10048.82, 10066.51, 10189.14, 10496.11~\AA\, and 
an unusual big scatter in abundances was noticed. Together with Fe and Mg, Ti overabundance was found in the atmosphere of Betelgeuse as well. 

The similar big scatter was noticed in Cr abundances, when we employed six Cr\ione\ lines at $\lambda\lambda$ 8707.402, 8707.965, 9016.98, 9035.85, 10486.25, 10510.01~\AA\,. 
However, using the same line list, the solar Cr abundance was obtained A(Cr) = 5.60$\pm$0.13 with similar big scatter. 
In contrast to Fe, Mg, and Ti, the mean values of Chromium in the atmosphere of Betelgeuse are found to be lower compared to solar Cr abundance. 

To measure Sr abundance, we used two strong Sr\ii\ lines at 10036.65~\AA\, and 10327.31~\AA\,. 
Solar Sr abundance from these lines is A(Sr) = 3.3$\pm$0.07, while in the atmosphere of Betelgeuse, Sr overabundance reaches up to [Sr/H]~$\sim$~1.4~dex.

\subsection*{ Approbation of the method.} 
Following the standard scientific process we have tested the method with additional 15 red supergiant stars of K- and M- spectral types by employing their high resolution spectra in the near-IR ranges.
The characteristics of the observed spectra are presented in {\bf Supplementary Table~6}, where $V$ magnitudes were adopted from SIMBAD data base, and S/N was measured on 8000~\AA\,.
The spectral types of HD~36389, $\alpha$~Her, HD~44537, HD~50877, HD~52877, HD~200905, HD~206936, HD~210745, and HD~216946 were adopted from ref. \cite{2005ApJ...628..973L}, and from SIMBAD data base for the remaining stars. The spectrum of HD~49331 was extracted from the ESO archive [\url{http://www.eso.org/sci/observing/tools/uvespop.html}]. 
For the remaining stars, their spectra have the same characteristics as the spectrum of Betelgeuse obtained on 2012 February 14. 

The physical parameters (\Teff, log~$g$, [Fe/H], and $V_{mic}$) derived in this study for 15 additional stars are presented in 
{\bf Table~2}. The obtained effective temperatures are ranged from 3440~K to 4139~K, and log~$g$, from $-$0.26 to +3.08. 
  
  The theoretical spectra were calculated with oxygen and titanium abundances, A(O) = 8.69 and A(Ti) = 4.95.  It should be noted that the values of oxygen and titanium abundances may affect on the absolute values of \Teff, which is determined on the base of TiO molecular lines. Although, we would not expect a deviation of Ti abundance from the solar value in the atmosphere of RSGs stars, while oxygen composition may deviate from the solar value due to more advanced evolutionary stage of RSGs. For example, if the oxygen abundance is higher  by 0.1 dex, the systematic shift $\sim$+40~K should be considered for the effective temperature. Other parameters, log~$g$, [Fe/H], and $V_{mic}$, were taken in accordance with 
 {\bf Table~2}. The theoretical spectra were convolved with a profile that combines a rotational broadening, broadening by macroturbulence and instrumental profile, $R$ = 65\,000. The macroturbulence was varied between 8 to 16 \kms\ for different stars.  
  The quality of the fits is illustrated in {\bf Figure~8} for four representing stars: $\alpha$~Her, VX Sgr, HD~217906, and HD~146051 with the effective temperatures in the range from 3440~K to 3729~K.  
   
The comparison with previous studies is presented in {\bf Supplementary Figure 5}. 
For \Teff\ less than 3800~K, our \Teff\ are in good agreement with data from ref. \cite{2005ApJ...628..973L}, where \Teff\ were derived from the MARCS models using broad-band photometry. 
For three stars with \Teff\ from $\sim$3800~K to $\sim$4100~K, our temperatures are higher by about $\sim$150~K compared to \cite{2005ApJ...628..973L} ({\bf Supplementary Figure 5a)}.

For nine common stars, our effective temperatures are in good agreement with ref. \cite{2012MNRAS.427..343M}, whose temperatures were obtained 
by comparing \textsc{bt-settl} model atmospheres to spectral energy distributions (SEDs) created from $Hipparcos$ data.
Although, there are three outliers. For VX Sgr, ref. \cite{2012MNRAS.427..343M} have obtained low temperature \Teff\ = 2255~K, while our result is \Teff\ = 3533~K. This star was not included in {\bf Supplementary Figure 5b}. Other two outliers are HD~52877 and HD~206936. For the first of them, our \Teff\ is lower by $\sim$200~K compared to ref. \cite{2012MNRAS.427..343M}, while for the second of them our \Teff\ is higher by $\sim$300~K compared to ref. \cite{2012MNRAS.427..343M}.

We found six common stars with ref. \cite{2011A&A...531A.165P}, who presented the stellar atmospheric parameters by employing \textsc{ULySS} program with the ELODIE spectral library. For three stars with \Teff\ $\le$ 3800~K, our temperatures are in line with ref. \cite{2011A&A...531A.165P}, however they are lower by $\sim$200~K for another 2 stars. One of the outliers is 32~Cyg (HD~192909). We obtained \Teff\ = 3771~K, which is almost in the middle between \Teff\ = 3543~K derived from interferometric observations by \cite{1998AJ....116..981D} and \Teff\ = 3978$\pm$46~K presented in ref. \cite{2011A&A...531A.165P}. For five common stars, our log~$g$ were found to be slightly lower by $\sim$0.40 dex compared to ref. \cite{2011A&A...531A.165P} {\bf (Supplementary Figure 5c, d)}, however still in a good agreement within 3$\sigma$.

We conclude that the presented method is sensitive to effective temperature and surface gravity and can be used for determining \Teff\ and log~$g$ for 
giants and supergiant stars of K- and M- spectral types based on their high resolution spectral observations in the near-IR spectral range.

\section*{Data availability}
All data generated and/or analysed during the current study are available from the corresponding author upon request. 
Source Data are provided with this paper.

\section*{Code availability}
Computer MCMC code is available from GitHub under \url{https://github.com/salex-asto/Betelgeuse-NC}. We made use of MARCS \cite{2008A&A...486..951G}, \textsc{synthV\_NLTE} \cite{2016MNRAS.456.1221R}, \textsc{detail} \cite{detail},  \textsc{emcee} package \cite{emcee}, \textsc{binmag} v.6 code [\url{http://www.astro.uu.se/~oleg/binmag.html}] \cite{binmag6}

\bibliography{Bet}

%\end{thebibliography}

\section*{Acknowledgements}

  This study is supported by the National Natural Science Foundation of China under grants Nos. 11988101, 11890694, 12050410265 and National Key R$\&$D Program of China No. 2019YFA0405502. This research is based on spectroscopic observations carried out at Weihai Observatory of Shandong University in People's Republic of China. S.A. acknowledges support from the LAMOST FELLOWSHIP that is budgeted and administrated by Chinese Academy of Sciences. This research used the facilities of the Canadian Astronomy Data Centre operated by the National Research Council of Canada with the support of the Canadian Space Agency. 
We acknowledge with thanks the variable star observations from the AAVSO International Database contributed by observers worldwide and used in this research.
We made use of NIST, SIMBAD and VALD databases. Based on observations collected at the European Southern Observatory under ESO programme 266.D-5655(A).

\section*{Author contributions statement}

G.Z. and S.A. proposed and designed this study. K.L. and S.H. provided observational time and organized the high-resolution spectroscopic observations. D.G. carried out the observations and made data reduction. A.L. participated in the discussion on the possible obscuring effects of dust. S.A. applied the theoretical methods and led the data analysis. J.D. developed MCMC Python code. All the authors discussed the results and contributed to the writing of the manuscript.

\section*{Competing Interests}

 The authors declare no competing interests.

\section*{Correspondence}

Correspondence and requests for materials should be addressed to Gang Zhao (gzhao@nao.cas.cn).

\newpage

  \begin{table*}
   \begin{center}   
   \caption{{\bf Physical parameters of Betelgeuse obtained in this study.} Julian date (JD-2450000), effective temperature (\Teff), surface gravity (log~$g$), metallicity ([Fe/H]), and microturbulent velocity ($V_{\rm mic}$)) are presented. Date of observations is given in dd-mm-yyyy format. Uncertainty for \Teff\ and log~$g$ ($\sigma$(\Teff) and $\sigma$(log~$g$)) is defined as the difference between the 84th and 50th percentile as the upper limit, and the difference between the 50th and 16th percentile as the lower limit. 
  Uncertainty for [Fe/H] ($\sigma$([Fe/H])) is calculated as standard deviation. Source data are provided as a Source Data file. }
   \label{tab_sum}
   \begin{tabular}{{|l|l|l|l|l|l|l|l|l|}}
   \hline
Date of observations & JD - 2450000 (d) & \Teff\ (K) & $\sigma$(\Teff) &  log~$g$  & $\sigma$(log~$g$)  &  [Fe/H]  &  $\sigma$([Fe/H]) &  $V_{\rm mic}$ (\kms) \\ \hline                                                                                       
06-04-2020 & 8945.6 & 3646 & 9 & 0.24& 0.04 & 0.11 & 0.07 & 3 \\ \hline                          
04-04-2020 & 8943.6 & 3611 & 7 & 0.17& 0.04 & 0.12 & 0.15 & 3 \\ \hline                          
19-03-2020 & 8927.6 & 3534 & 3 & 0.25& 0.04 & 0.06 & 0.09 & 3 \\ \hline                          
31-01-2020 & 8879.6 & 3476 & 4 & 0.22& 0.04 & 0.12 & 0.12 &  3 \\  \hline                         
14-02-2012 & 6702.6 & 3632 & 7 & 0.11& 0.03 & 0.07 & 0.16 & 3 \\\hline     
    \end{tabular}
  \end{center}
\end{table*}

\clearpage

  \begin{table*}
   \begin{center}   
   \caption{{\bf Physical parameters of red supergiants obtained in this study.} Effective temperature (\Teff), surface gravity (log~$g$), metallicity ([Fe/H]), and microturbulent velocity ($V_{mic}$) are given for each star. 
   Uncertainties for \Teff\ (-$\sigma$(\Teff) and +$\sigma$(\Teff)) and log~$g$ (-$\sigma$(log~$g$) and +$\sigma$(log~$g$)) are defined as the difference between the 84th and 50th percentile as the upper limit, and the difference between the 50th and 16th percentile as the lower limit. 
  Uncertainties for [Fe/H] ($\sigma$([Fe/H])) are calculated as standard deviation. Source data are provided as a Source Data file. }
   \setlength{\tabcolsep}{2pt} 
   \label{tab_res_others}
\begin{tabular}{|l|l|l|l|l|l|l|l|l|l|l|}\hline
   Star   &  Name        & \Teff\ (K) & -$\sigma$(\Teff) & +$\sigma$(\Teff)   & log~$g$ & -$\sigma$(log~$g$) & +$\sigma$(log~$g$) & [Fe/H]    & $\sigma$([Fe/H]) & $V_{mic}$ (\kms)  \\    \hline                                                                                      
HD~36389  & 119 Tau      & 3610       &  -8              &  +8                & ~0.02   &  -0.01  &    +0.02                      & ~0.03     & 0.09  & 2.5   \\\hline
HD~44537  & psi01 Aur    & 3723       &  -9              &  +9                &$-$0.16  &  -0.08  &    +0.08                      & ~0.07     & 0.10  & 3     \\\hline
HD~49331  & --           & 3691       &  -5              &  +4                & ~0.41   &  -0.02  &    +0.02                      & $-$0.23   & 0.12  & 3     \\\hline
HD~50877  & omi01 CMa    & 4095       &-20               &  +21               &$-$0.15  &  -0.10  &    +0.08                      & $-$0.02   & 0.19  & 3     \\\hline
HD~52877  & $\sigma$~CMa & 3735       &  -4              &  +4                & ~0.22   &  -0.08  &    +0.08                      & ~0.06     & 0.12  & 2     \\\hline
HD~146051 & $\delta$~Oph & 3729       &  -8              &  +8                & ~0.98   &  -0.02  &    +0.01                      & $-$0.38   & 0.05  & 2     \\\hline
HD~156014 & $\alpha$~Her & 3440       &  -5              &  +5                & ~3.08   &  -0.15  &    +0.15                      & ~0.59     & 0.06  & 3     \\\hline
HD~165674 & VX Sgr       & 3533       &-10               &  +11               & ~1.98   &  -0.03  &    +0.02                      & ~0.61     & 0.45  & 3     \\  \hline
HD~192909 & 32 Cyg       & 3771       &  -8              &  +8                & ~0.45   &  -0.02  &    +0.02                      & $-$0.03   & 0.11  & 2     \\\hline
HD~200905 & $\xi$~Cyg    & 3964       &-14               &  +14               & ~0.76   &  -0.02  &    +0.02                      & $-$0.01   & 0.15  & 2     \\\hline
HD~206936 & $\mu$~Cep    & 3683       &  -9              &  +9                &$-$0.26  &  -0.08  &    +0.08                      & ~0.26     & 0.25  & 3     \\\hline
HD~208816 & VV Cep       & 3662       &-10               &  +10               & ~0.09   &  -0.02  &    +0.02                      & ~0.10     & 0.04  & 3     \\  \hline
HD~210745 & $\zeta$~Cep  & 4139       &-112              &  +77               & ~0.59   &  -0.03  &    +0.03                      & ~0.22     & 0.18  & 2     \\\hline
HD~216946 & V424 Lac     & 3815       &-74               &  +93               & ~0.49   &  -0.03  &    +0.03                      & ~0.21     & 0.10  & 2     \\\hline
HD~217906 & $\beta$~Peg  & 3621       & -5               &  +5                & ~1.15   &  -0.05  &    +0.05                      & $-$0.22   & 0.04  & 2     \\\hline   
    \end{tabular}
  \end{center}
\end{table*}

\begin{figure*}
%\begin{minipage}{160mm}
%\parbox{0.99\linewidth}{
\includegraphics[scale=0.5]{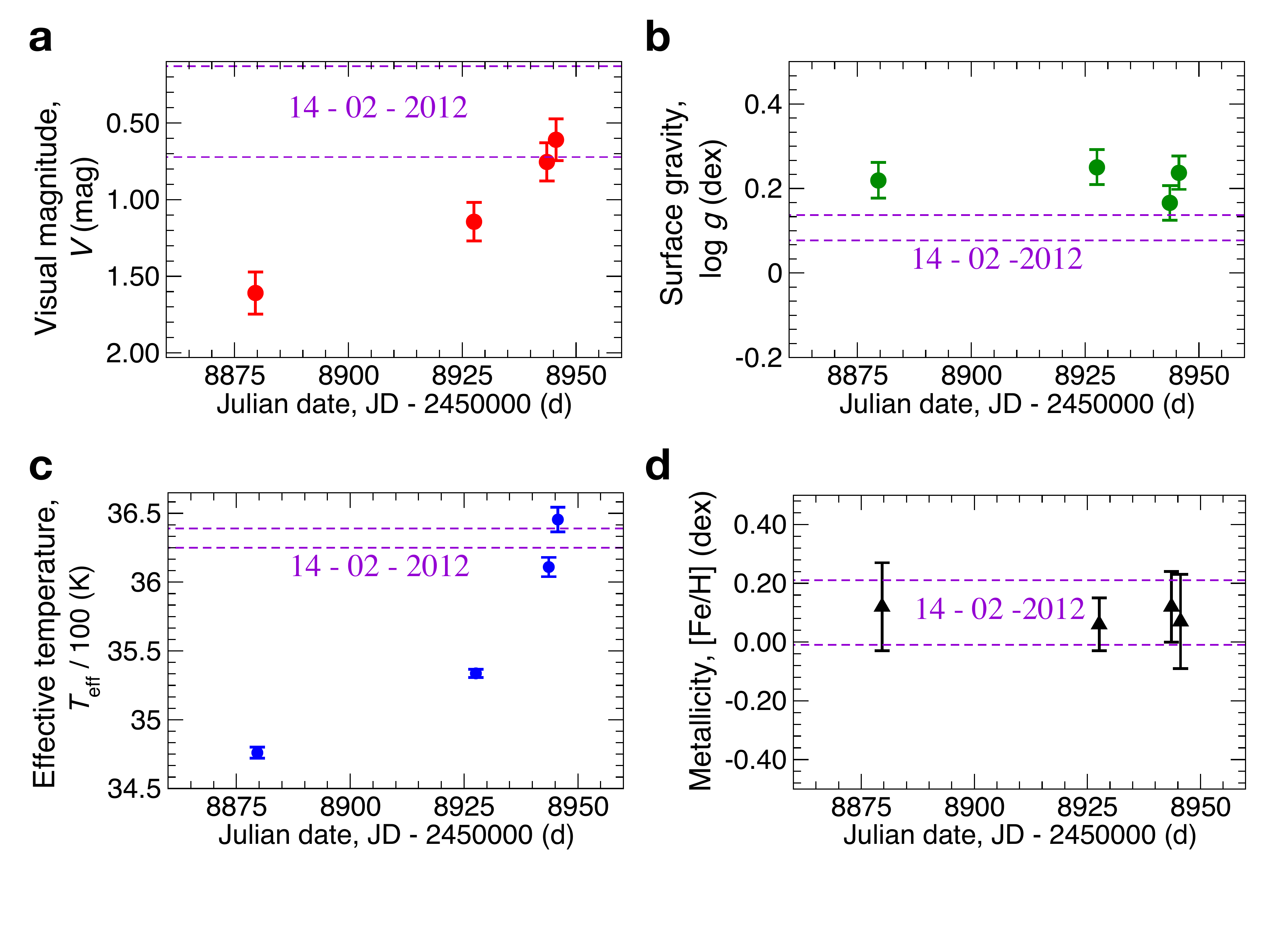}\\
%\centering}
\hfill
\\[0ex]
\caption{{\bf The mean visual magnitude ($V$), surface gravity (log~$g$), effective temperature (\Teff ), and metallicity ([Fe/H]) of Betelgeuse obtained on 2020 January 31, March 19, April 4, and April 6. }
{\bf a} Each V magnitude was calculated by averaging over all available magnitudes in the AAVSO database in a particular date of observations. 
{\bf b} The uncertainty of surface gravity (log~$g$) is taken as the difference between the 84th and 50th percentile as the upper limit, and the difference between the 50th and 16th percentile as the lower limit. 
{\bf c} The uncertainty of the effective temperature (\Teff) is taken as the difference between the 84th and 50th percentile as the upper limit, and the difference between the 50th and 16th percentile as the lower limit. 
{\bf d} Metallicity ([Fe/H]) was calculated by averaging over all iron abundances from individual lines in a particular date of observations. The two horizontal dashed lines show the ranges of parameters ($V$, \Teff , log~$g$, [Fe/H]) for pre-dimming, which we have obtained with ESPaDOnS's spectrum observed on 2012 February, 14. Error bars represent standard deviation. Source data are provided as a Source Data file.}
\label{Figure1}
%\end{minipage}
\end{figure*}

\begin{figure*}
%\begin{minipage}{160mm}
\parbox{0.99\linewidth}{
\includegraphics[scale=0.6]{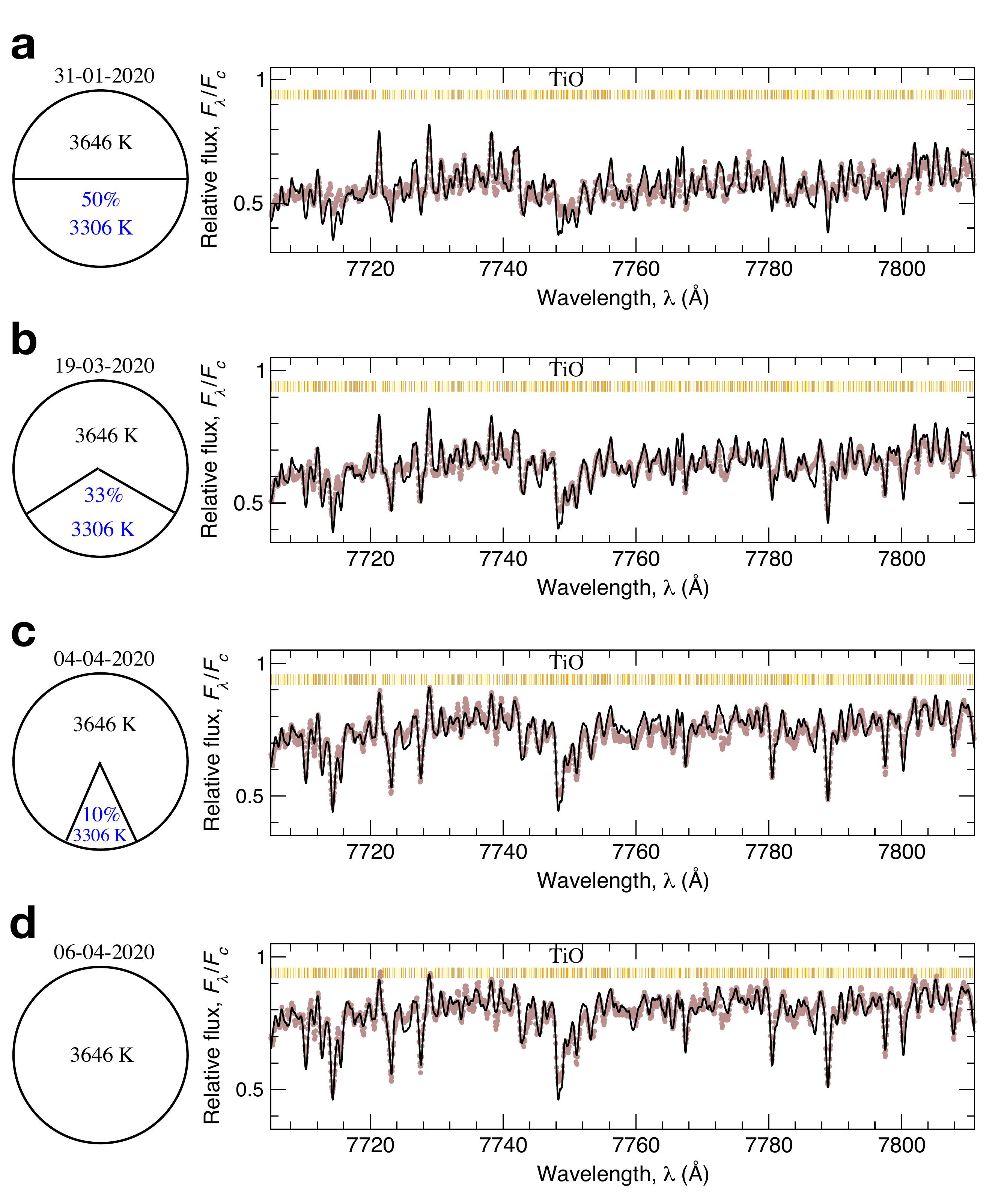}\\
\centering}
\hfill
\\[0ex]
\caption{{\bf The schemes of stellar surface with a dark spot and the corresponding best fits of the narrow spectral region 7705 -- 7811~\AA\, for four dates of observations.} 
The temperature increasing is interpreted as the changing of surface area of dark spot. Relative flux is calculated as a ration of the flux at particular wavelength ($F_\lambda$) to the flux in continuum ($F_c$). The observed spectra are shown as brown dots. The theoretical spectra (black solid curves) were calculated with the obtained parameters from {\bf Table~1} and convolved with the instrumental profile. Everywhere $V_{mic}$ = 3 \kms, $V$~sin$i$ = 5 \kms, $V_{mac}$ = 16 \kms. Titanium, carbon, nitrogen, and oxygen abundances are taken to be 4.95, 8.43, 8.60, and 8.80 dex, respectively.  The molecular TiO (titanium oxide) lines are marked by vertical lines on each panel. 
{\bf a} On 31-01-2020, the temperature of spot is 3306~K if it occupies 50$\%$ of surface area. {\bf b} On 19-03-2020, the temperature of spot is 3306~K if it occupies 33$\%$ of surface area. {\bf c} On 04-04-2020, the temperature of spot is 3306~K if it occupies 10$\%$ of surface area.  
{\bf d} On 06-04-2020, the temperature of whole star is 3646~K.
Source data are provided as a Source Data file. }
\label{Figure2}
%\end{minipage}
\end{figure*}

\begin{figure*}
%\begin{minipage}{160mm}
\parbox{0.99\linewidth}{
\includegraphics[scale=0.5]{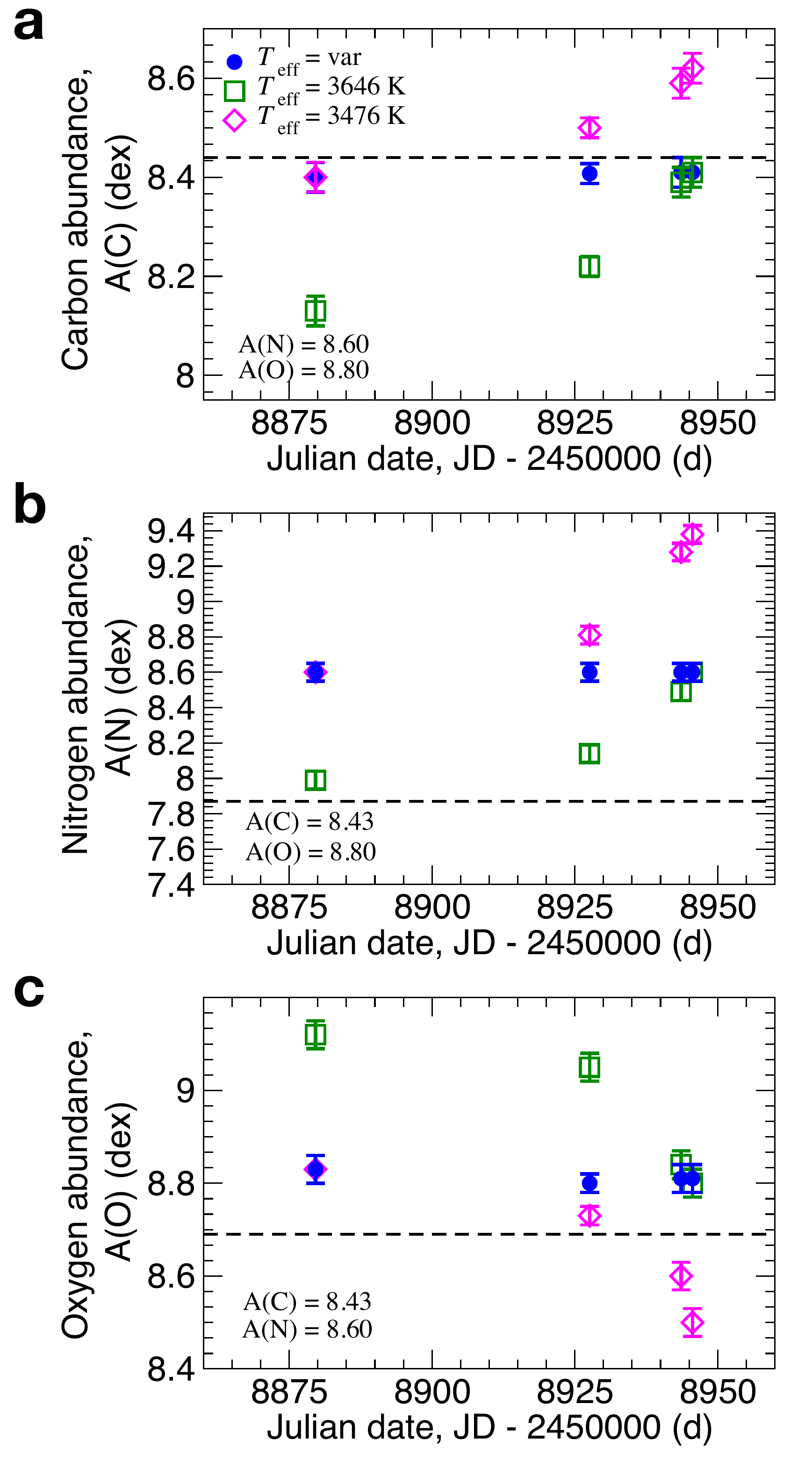}\\
\centering}
\hfill
\\[0ex]
\caption{ {\bf Conditional chemical abundances of carbon, nitrogen, and oxygen in the atmosphere of Betelgeuse at four different epochs: 31-01-2020, 19-03-2020, 04-04-2020, and 06-04-2020. } 
Chemical abundances of carbon, nitrogen, and oxygen are calculated under three scenarios. First scenario is effective temperature \Teff\ = var, the chemical abundances are obtained at parameters according to {\bf Table 1} (blue filled circles). Second scenario is the chemical abundances are obtained at the assumption of constant effective temperature \Teff\ = 3646~K and log~$g$ = 0.2 (green open squares). 
Third scenario is the chemical abundances are obtained at the assumption of constant effective temperature \Teff\ = 3476~K and log~$g$ = 0.2 (pink open diamonds). 
{\bf a} Carbon abundances calculated under three scenarios. 
{\bf b} Nitrogen abundances calculated under three scenarios.  
{\bf c} Oxygen abundances calculated under three scenarios.  
The dashed lines show solar abundances obtained in this study. Error bars represent standard deviation. 
 Source data are provided as a Source Data file.}
\label{Figure3}
%\end{minipage}
\end{figure*}

\begin{figure*}
%\begin{minipage}{160mm}
\parbox{0.99\linewidth}{
\includegraphics[scale=0.28]{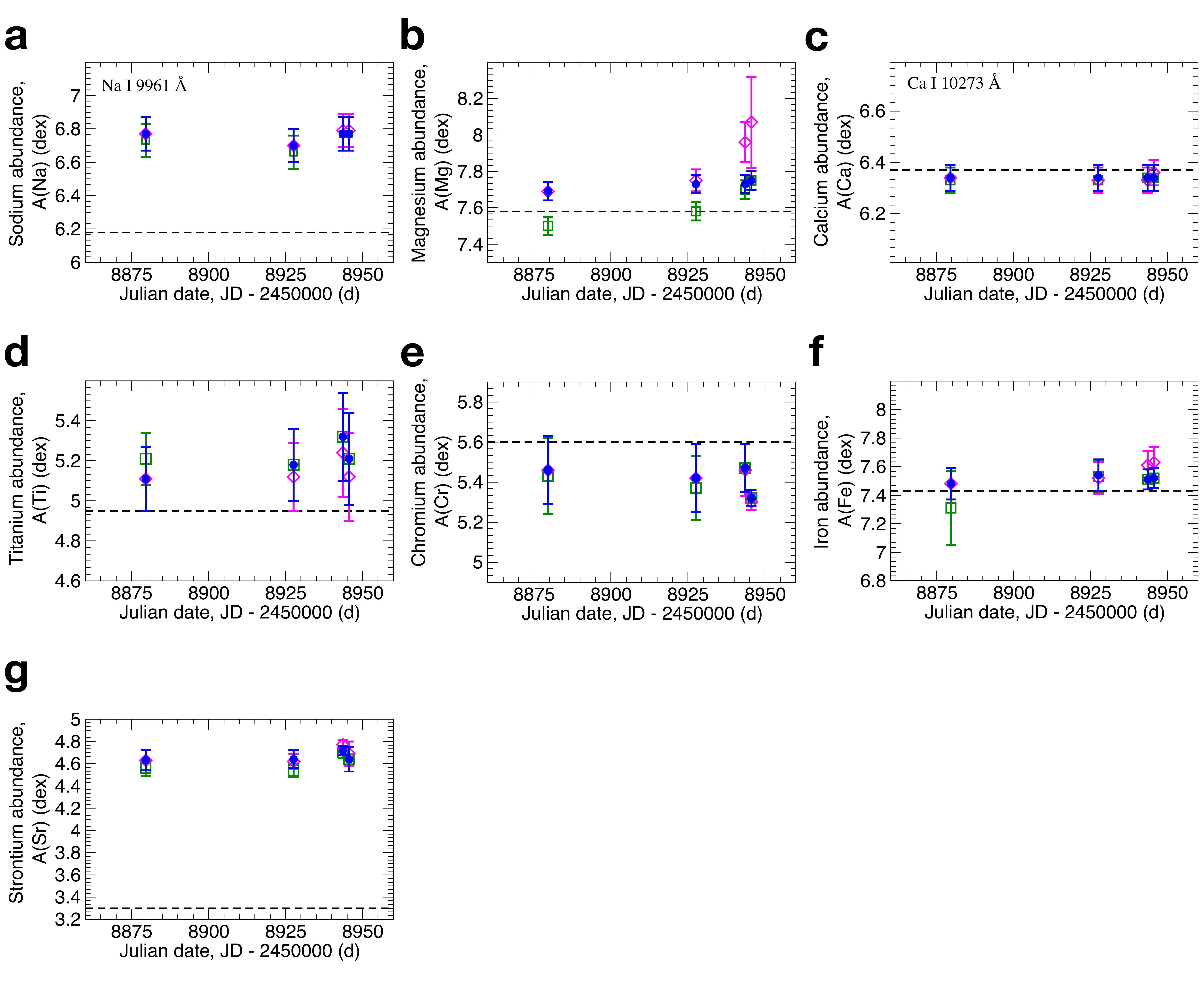}\\
\centering}
\hfill
\\[0ex]
\caption{ {\bf Chemical composition in the atmosphere of Betelgeuse at four different epochs: 31-01-2020, 19-03-2020, 04-04-2020, and 06-04-2020.  } 
{\bf a} Na abundances calculated under three scenarios. 
{\bf b} Mg abundances calculated under three scenarios.  
{\bf c} Ca abundances calculated under three scenarios.  
{\bf d} Ti abundances calculated under three scenarios. 
{\bf e} Cr abundances calculated under three scenarios. 
{\bf f} Fe abundances calculated under three scenarios.  
{\bf g} Sr abundances calculated under three scenarios.  
Everywhere, the meanings of symbols are the same as in {\bf Figure 3}. The dashed lines show solar abundances obtained in this study. Error bars represent standard deviation. Source data are provided as a Source Data file.}
\label{Figure4}
%\end{minipage}
\end{figure*}

\begin{figure*}
%\begin{minipage}{160mm}
\parbox{0.99\linewidth}{
\includegraphics[scale=0.5]{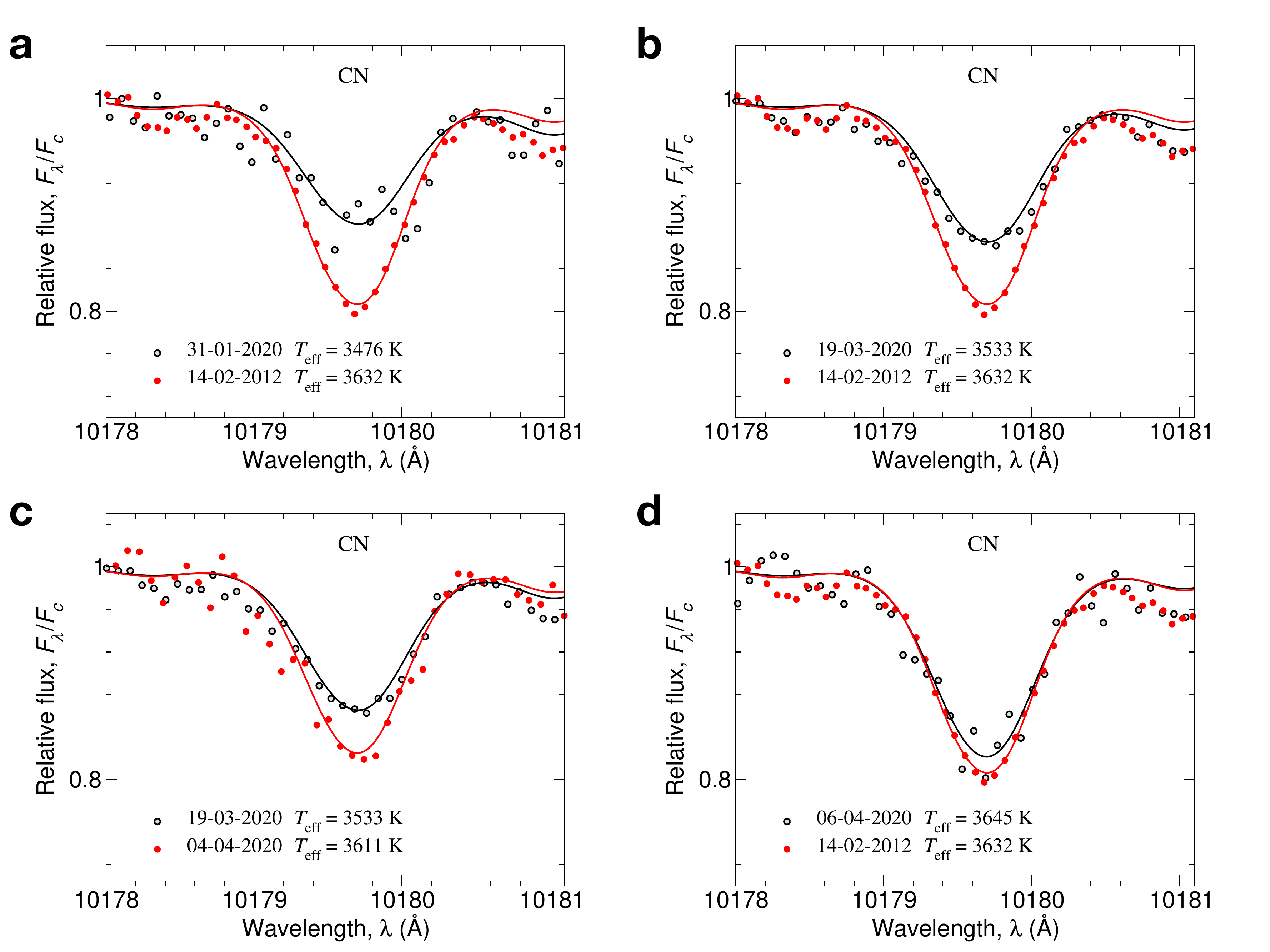}\\
\centering}
\hfill
\\[0ex]
\caption{ {\bf The different strength of CN (cyanide) molecular line at 10179~\AA\  at different epochs can be explained by different temperatures at these epochs. } 
Relative flux is calculated as a ration of the flux at particular wavelength ($F_\lambda$) to the flux in continuum ($F_c$). The spectral observations are shown by dots. Solid curves are theoretical spectra calculated at the particular temperature specified on each panel, convolved with the instrumental profile. Carbon, nitrogen, and oxygen abundances are 8.43, 8.60, and 8.80 dex, respectively. This line was presented as an example. The similar behavior was found for other CN lines at $\lambda\lambda$ 10223, 10367, 10376, and 10377~\AA.
{\bf a} The CN molecular line became stronger on 14-02-2012 compared to 31-01-2020. On the panel {\bf b}, the comparison of CN lines observed on 14-02-2012 and 19-03-2020. 
{\bf c} This CN molecular line is stronger on 19-03-2020 compared to 04-04-2020, while CN molecular lines look similar at both epochs 14-02-2012 and 06-04-2020 for which the temperature difference is only 13 K 
(see {\bf d}). Source data are provided as a Source Data file.}
\label{Figure5}
%\end{minipage}
\end{figure*}

\begin{figure*}
\begin{minipage}{180mm}
\parbox{0.99\linewidth}{
\includegraphics[scale=0.5]{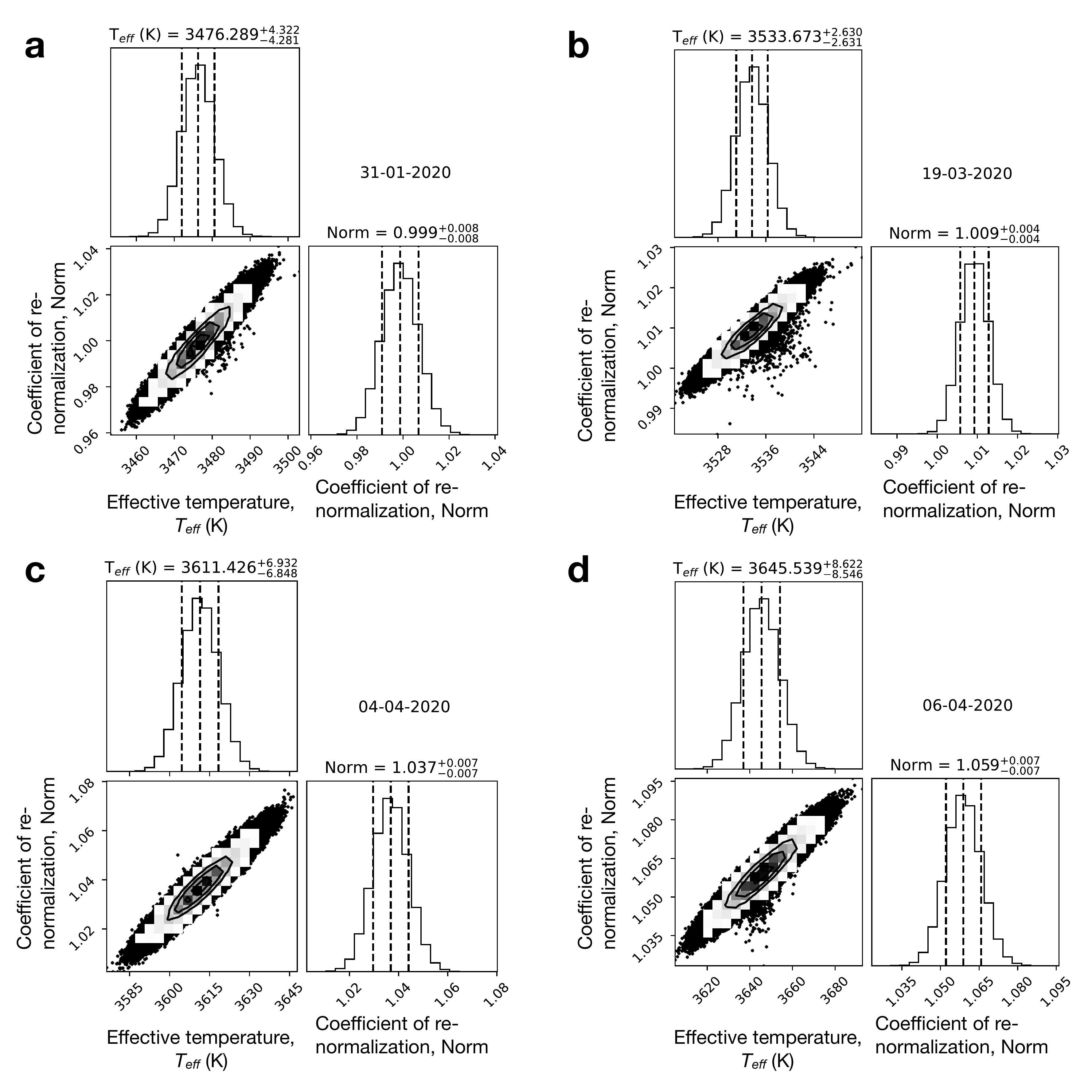}\\
\centering}
\hfill
\\[0ex]
\caption{  {\bf Four corner plots with the final effective temperatures (\Teff) of Betelgeuse obtained at four epochs 31-01-2020, 19-03-2020, 04-04-2020, and 06-04-2020.}  
{\bf a} The best-fit \Teff\ and its uncertainties obtained on 31-01-2020. The panel corresponding to the fit of the region at 7700 -- 7900 \AA\ shown in {\bf Supplementary Figure 4a}. 
{\bf b} The best-fit \Teff\ and its uncertainties obtained on 19-03-2020. The panel corresponding to the fit of the region at 7700 --  7900 \AA\ shown in {\bf Supplementary Figure 4b}. 
{\bf c} The best-fit \Teff\ and its uncertainties obtained on 04-04-2020. The panel corresponding to the fit of the region at 7700 -- 7900 \AA\ shown in {\bf Supplementary Figure 4c}. 
{\bf d} The best-fit \Teff\ and its uncertainties obtained on 06-04-2020. The panel corresponding to the fit of the region at 7700 -- 7900 \AA\ shown in {\bf Supplementary Figure 4d}. 
On each panel, the diagonal shows the marginalized posteriors. The subsequent covariances between all the parameters are in the corresponding 2D histograms. 
The vertical lines represent the 16, 50 and 84 percentiles. The best-fit parameters for \Teff\ and their uncertainties are presented on the top of each panel. Uncertainty for \Teff\ is defined as the difference between the 84th and 50th percentile as the upper limit, and the difference between the 50th and 16th percentile as the lower limit. The data of observations are marked in dd-mm-yyyy format.}
\label{Figure6}
\end{minipage}
\end{figure*}

\begin{figure*}
\begin{minipage}{180mm}
\parbox{0.99\linewidth}{
\includegraphics[scale=0.25]{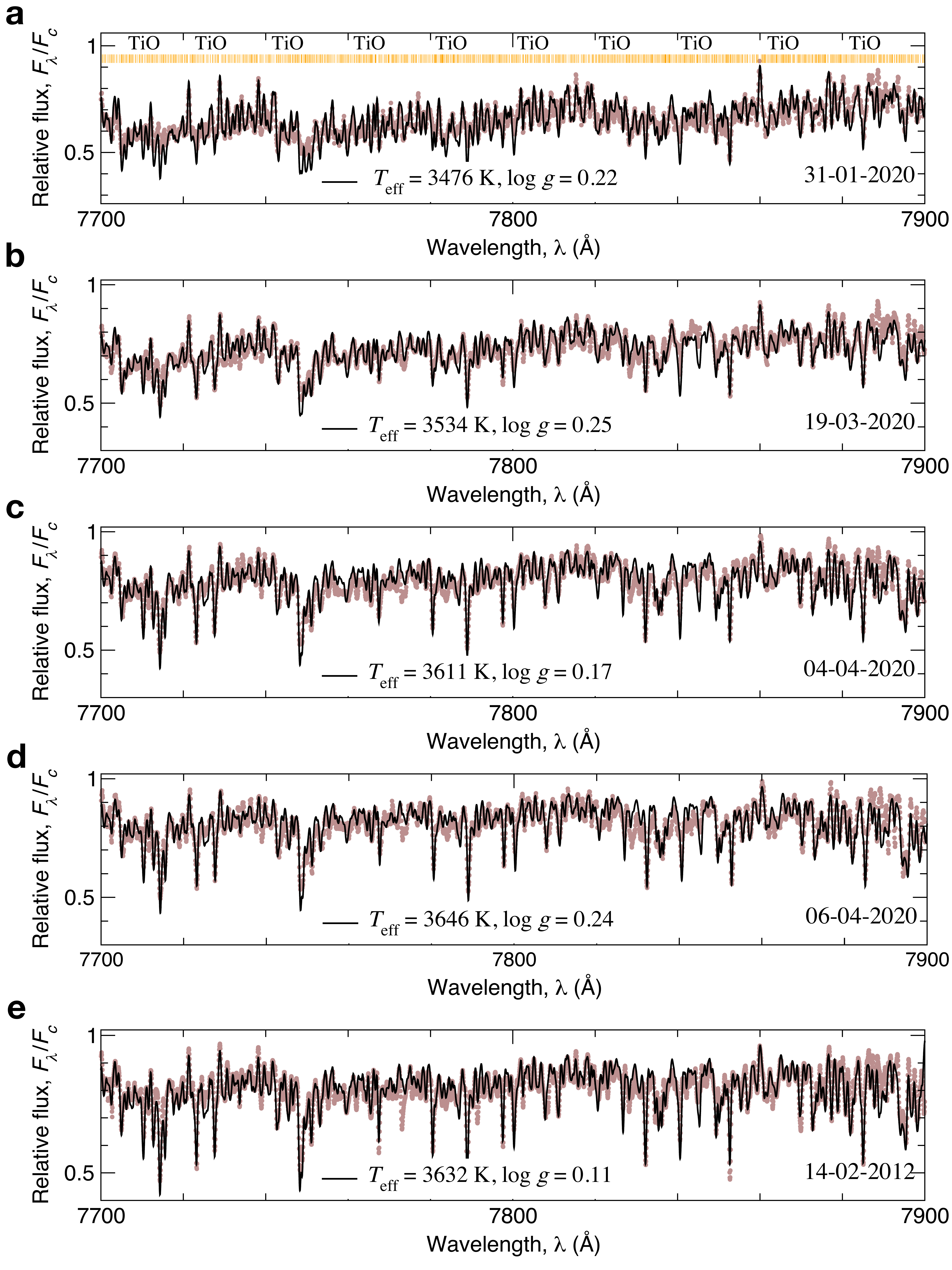}\\
\centering}
\hfill
\\[0ex]
     \caption{ {\bf Best fits to the molecular bands of TiO for five sets of spectra.} The observed spectra of Betelgeuse obtained on 31-01-2020 ({\bf a}), 19-03-2020 ({\bf b}), 04-04-2020 ({\bf c}), 06-04-2020 ({\bf d}), and 14-02-2012 ({\bf e}) are shown by brown dots. Solid line is theoretical profile calculated with the particular effective temperature (\Teff) and surface gravity (log~$g$), which were obtained in this study and presented on each panel. Relative flux is calculated as a ration of the flux at particular wavelength ($F_\lambda$) to the flux in continuum ($F_c$). The 2365 TiO lines are marked by vertical lines on the panel ({\bf a}). Source data are provided as a Source Data file.}
\label{Figure7}
\end{minipage}
\end{figure*}

\begin{figure*}
%\begin{minipage}{160mm}
\parbox{0.99\linewidth}{
\includegraphics[scale=0.49]{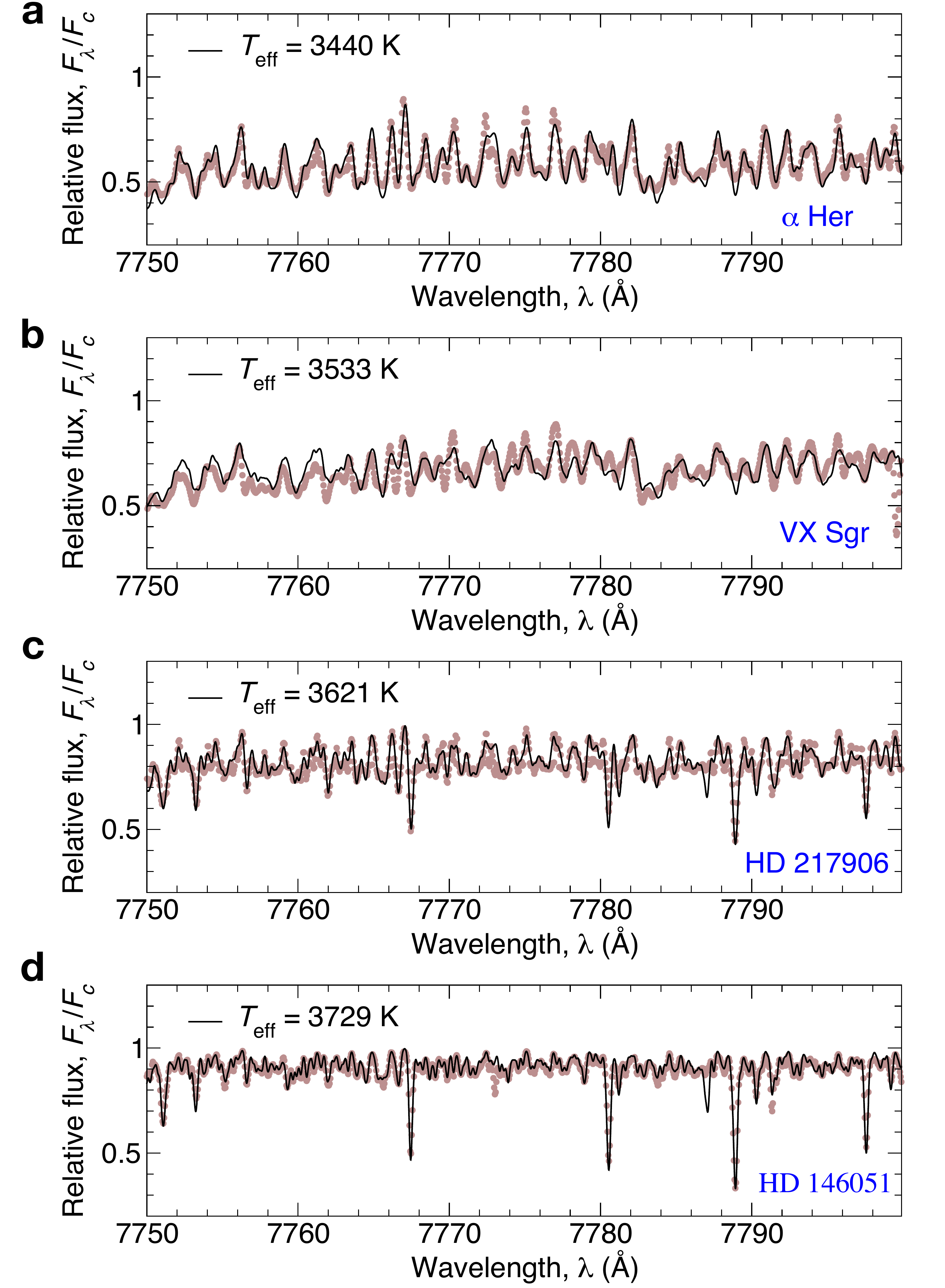}\\
\centering}
\hfill
\\[0ex]
\caption{{\bf The best fits of the parts of spectra of Betelgeuse in 7750--7800~\AA\ range for four represented supergiant stars.} Relative flux is calculated as a ration of the flux at particular wavelength ($F_\lambda$) to the flux in continuum ($F_c$). The solid black line is theoretical profile calculated with the particular \Teff\ obtained in this study. The observed spectra are shown by brown dots. The name of the star and its \Teff\ are presented on each panel. {\bf a} The best fit for the coolest star in our sample, $\alpha$~Her, with effective temperature (\Teff) 3440~K.
{\bf b} The best fit for VX Sgr with \Teff\ = 3533~K.
{\bf c} The best fit for HD~217906 with \Teff\ = 3621~K. 
{\bf d} The best fit for HD~146051 with \Teff\ = 3729~K. 
Source data are provided as a Source Data file. }
\label{Figure6}
%\end{minipage}
\end{figure*}

\end{document}